\documentclass{kapproc} 

\let\footnote\savefootnote

\usepackage{amsmath}
\usepackage{graphicx}%
\usepackage{amsfonts}%
\usepackage{amssymb}
\usepackage{procps}

\upperandlowercase 
\sectionequationnumber 

\normallatexbib

\def\nin{\mbox{$\not\!\,\in\,$}}

\newcommand{\R} {\mbox{Re}\,}

\newcommand{\la}{\label}
\newcommand{\be}{\begin{equation}}
\newcommand{\ee}{\end{equation}}
\newcommand{\bea}{\begin{eqnarray}}
\newcommand{\eea}{\end{eqnarray}}

\begin{document}
\articletitle{Hydrodynamics of correlated systems}
\articlesubtitle{Emptiness Formation Probability and Random Matrices\thanks{This is an extended version of the seminar given at the School  "Applications of Random Matrices in Physics", Les Houches, June 2004}}
\author{Alexander G. Abanov}
\affil{Department of Physics and Astronomy, 
Stony Brook University,  \\ Stony Brook, NY 11794-3800, U.S.A.}
\email{alexandre.abanov@sunysb.edu}



\begin{abstract}
A hydrodynamic approach is used to calculate an asymptotics of the \textit{Emptiness Formation Probability} -- the probability of a formation of an empty space in the ground state of a quantum one-dimensional many body system. Quantum hydrodynamics of a system is represented as a Euclidian path integral over configurations of hydrodynamic variables. In the limit of a large size of the empty space, the probability is dominated by an instanton configuration, and the problem is reduced to the finding of an instanton solution of classical hydrodynamic equations. After establishing a general formalism, we carry out this calculation for several simple systems -- free fermions with an arbitrary dispersion and Calogero-Sutherland model. For these systems we confirm the obtained results by comparison with exact results known in Random Matrix theory. We argue that the nonlinear hydrodynamic approach might be useful even in cases where the linearized hydrodynamics fails.

\end{abstract}

\begin{keywords}
Quantum Hydrodynamics, Random Matrices, Instanton, Rare Fluctuation, Emptiness Formation Probability, Calogero-Sutherland Model.
\end{keywords}

\newpage
\tableofcontents
\setcounter{page}{3}

\chaptitlerunninghead{Hydrodynamics of correlated systems \ldots}


\section{Introduction}
\label{sec_intro}

The hydrodynamic approach to correlation functions in quantum many body systems has a long history \cite{Landau-1941, Forster-1975}.  Generally, hydrodynamic equations are nonlinear and dispersive. However, usually, it is the linearized hydrodynamics which is used to extract long distance asymptotics of correlation functions \cite{Forster-1975}. In particular, in one spatial dimension the linearized quantum hydrodynamics (or bosonization) is especially useful \cite{Stone-book}. It was argued that corrections to the bosonization due to the nonlinearities of the hydrodynamics (curvature of an underlying fermionic spectrum) are irrelevant in calculation of leading terms of correlation functions \cite{Haldane-1981}.

Although the linearized hydrodynamics turned out to be a very powerful tool in studies of correlation functions, there are many phenomena for which nonlinearities are essential. The goal of this lecture is to consider a particular example where the linear approximation fails but nonlinear hydrodynamics can still be used to extract non-perturbative results for correlation functions. 

As such an example we consider a particular quantity which plays an important role in the theory of quantum, one-dimensional, integrable systems \cite{korepin93} -- the Emptiness Formation Probability (EFP). This correlation function was argued to be the simplest of correlators in some integrable models \cite{korepin93}. The EFP $P(R)$ is essentially, a probability of formation of an empty region of a size $2R$ in the ground state of the many body system. In integrable models the EFP has an exact representation in terms of determinants of Fredholm operators \cite{korepin93,korepin94} or multiple integrals \cite{jimbo}. These expressions are exact but are very complex, and extracting, e.g., long distance asymptotics $R\to \infty$ from the exact expressions is a non-trivial problem (see Appendix A and references therein for some known results). However, the limit of large $R$ is precisely the limit where hydrodynamic description is applicable, and we illustrate that it, indeed, correctly produces the leading term of the asymptotics of the EFP. The hydrodynamic approach can also be used to calculate more complicated quantities than EFP (which is, essentially, not a dynamic but a ground state property). Moreover, the applicability of hydrodynamics is not limited to integrable systems.   

We start with some definitions as well as with making a connection to Random Matrices.

\paragraph{Emptiness Formation Probability}

Consider a one-dimensional quantum liquid at zero temperature. The wave function of the ground state of the liquid $\Psi_{G}(x_{1}, x_{2},\ldots, x_{N})$ gives the probability distribution $|\Psi_{G}|^{2}$ of having all $N$ particles at given positions $x_{j}$, where $j=1,\ldots, N$. We introduce the \textit{Emptiness Formation Probability}  (EFP) $P(R)$ as a probability of having no particles with coordinates $-R<x_{j}<R$. Formally we define
\be
	P(R) = \frac{1}{\langle\Psi_{G}|\Psi_{G}\rangle}\int_{|x_{j}|>R}dx_{1}
	\ldots dx_{N}\; |\Psi_{G}(x_{1},\ldots,x_{N})|^{2},
 \la{efpdef}
\ee
or following Ref.\cite{korepin93}
\be 
	P(R) = \lim_{\alpha\to +\infty}\left\langle \Psi_{G}\right|
	e^{-\alpha \int_{-R}^{R}\rho(x)\,dx}\left|\Psi_{G}\right\rangle,
\ee
where $\rho(x)$ is a particle density operator
\be
	\rho(x) \equiv \sum_{j=1}^{N}\delta(x-x_{j}).
\ee
We are interested in an asymptotic behavior of $P(R)$ as $\rho_{0}R\to \infty$, where $\rho_{0}$ is the average density of particles in the system. EFP $P(R)$ gives us the probability that the one-dimensional river parts to make a ford of a macroscopic size $2R$.

\paragraph{Random Matrices}
\label{sec_random}

The EFP (\ref{efpdef}) introduced for a general one-dimensional quantum liquid is a well-known quantity in the context of spectra of random matrices \cite{dysmehta}. Namely, it is the probability of having no eigenvalues in some range. Consider e.g., the joint eigenvalue distribution for the Circular Unitary Ensemble (CUE). The CUE is defined as an ensemble of $N\times N$ unitary matrices with the measure given by de Haar measure. Diagonalizing matrices and integrating out unitary rotations, one obtains \cite{mehta}
\be
	\int DU \to \int \prod_{j=1}^{N}d\theta_{j}\, \prod_{1\le j<k\le N}
	\left|e^{i\theta_{j}}-e^{i\theta_{k}}\right|^{\beta},
\ee
where $\beta = 2$ for CUE and $e^{i\theta_{j}}$ with $j=1,\ldots,N$ are the eigenvalues of a unitary matrix. One can read the joint eigenvalue distribution
\bea
	P_{N}(\theta_{1},\ldots,\theta_{N}) &=& const. \prod_{1\le j<k \le N}
	\left|e^{i\theta_{j}}-e^{i\theta_{k}}\right|^{\beta}
 \la{jed} \\
	&\sim &
	\exp\left\{\frac{\beta}{2}N^{2} \int \frac{d\theta\,d\theta'}{(2\pi)^{2}}\,
	\rho(\theta)\ln|e^{i\theta_{j}}-e^{i\theta_{k}}| \rho(\theta')\right\}.
 \nonumber
\eea
Here we replaced the sums over particles (eigenvalues) to integrals with particle densities. We left only terms which are dominant in the limit of large $N$. \footnote{In fact, one can do a better job including subdominant corrections. See Ref.\cite{mehta} for details.} Now we introduce the probability of having no eigenvalues on the arc $-\alpha<\theta<\alpha$ as
\be
	P(\alpha) = \frac{1}{\cal N} \int_{\theta_{j}\nin [-\alpha,\alpha]}
	\prod_{j=1}^{N}d\theta_{j}\, \prod_{1\leq j<k\leq N}\left|
	e^{i\theta_{j}}-e^{i\theta_{k}}\right|^{\beta}.
\ee
This quantity is known as $E_{\beta}(0,\alpha)$ in notations of \cite{mehta}. For orthogonal, unitary, and symplectic circular ensembles the joint eigenvalue distribution is given by (\ref{jed}) with $\beta=1,2,4$ respectively.

\paragraph{Spin chains and lattice fermions}
\label{sec_spin}

The EFP can also be defined for spin chains where we are interested in the probability of having a ferromagnetic string of the length $n$ in the ground state of the spin chain. The Jordan-Wigner transformation maps spin $1/2$ chain to a one-dimensional lattice gas of spinless fermions. Under this mapping the ferromagnetic string corresponds to a string of empty lattice sites and one can write
\be
	P(R) =\left\langle \prod_{j=-R}^{R}\psi_{j}\psi_{j}^{\dagger} \right\rangle,
\ee
where $\psi_{j},\psi_{j}^{\dagger}$ are annihilation and creation operators of spinless fermions on the lattice site $j$.
Therefore, the probability of the formation of a ferromagnetic string in spin chains corresponds to the Emptiness Formation Probability of Jordan-Wigner fermions. We are  going to use a  language of particles in this paper but all results are also valid for corresponding one-dimensional spin systems.


\section{Instanton or rare fluctuation method}
\label{sec_instanton}

In the limit of large $R$ ($\rho_{0}R\gg 1$) we use a collective description instead of dealing with individual particles. We assume that $u$ is some collective field describing the hydrodynamic motion of a one-dimensional liquid. \footnote{Later we will use the conventional displacement field as $u$. See Eq. (\ref{displacement}).} The dynamics of the liquid is defined by a Euclidian partition function
\be
	Z = \int Du\; e^{-S[u]},
 \la{Z}
\ee
where $S[u]$ is the Euclidian action
\be
	S[u] = \int dx\,\int_{-1/2T}^{1/2T}d\tau\, L(u,\dot u)
 \label{act}
\ee
and the inverse temperature $1/T$ defines periodic boundary conditions in the imaginary time $\tau$. 

The asymptotic behavior of $P(R)$ in the limit of large $R$ is defined by a rare fluctuation when all particles move away so that at some time $t=0$ we have no particles in  the spatial interval $\left[-R,R\right]$. Then with an exponential accuracy
\be
	P(R) \sim e^{-S_{opt}}.
 \label{prinst}
\ee
Here $S_{opt}$ is the value of the action (\ref{act}) on the trajectory $u(x,t)$ which minimizes (\ref{act}) and is subject to EFP boundary conditions. These are: the EFP boundary condition
\be
	\rho(t=0;-R<x<R) =0,
 \la{bcefp}
\ee
and standard boundary conditions at infinity
\bea
	\rho &\to& \rho_{0}, \;\;\;\; x,\tau\to\infty,
 \nonumber \\
 	v &\to& 0, \;\;\;\; x,\tau\to\infty,
 \la{bcinf}
\eea
where $\rho,v$ are the density and velocity of particles related somehow to the collective field $u$. 

Let us estimate the probability of such rare fluctuation\footnote{In this section we follow closely the qualitative argument of Ref.\cite{abanovkor}}. Without the boundary condition (\ref{bcefp}) the minimum of the action subject to (\ref{bcinf}) is obviously given by the constant solution $\rho(x,t)=\rho_{0}$, $v(x,t)=0$. The condition (\ref{bcefp}), however, disturbs this constant solution in some area of space-time around the origin. The spatial extent of this disturbance is of the order of $R$. If we assume that the quantum system we are dealing with is some \textit{compressible} liquid we expect that the typical temporal scale of the disturbance is of the order of $R/v_{s}$, where $v_{s}$ is a sound velocity  at $\rho=\rho_{0}$. We conclude that the (space-time) ``area'' of the disturbance scales as $R^{2}$ and the action $S_{opt}\sim R^{2}$. Therefore, we expect a Gaussian decay for (\ref{prinst})
\be
	P(R) \sim e^{-\alpha R^{2}},
 \la{gauss}
\ee
where $\alpha$ is some (non-universal) constant depending on the details of (\ref{act}).

This argument can be extended for the case of low but finite temperature. Namely, while the temporal extent of the instanton $R/v_{s}$ is smaller than the inverse temperature $R/v_{s}\ll 1/T$ instanton ``does not know'' that the temperature is not zero and one obtains an intermediate Gaussian decay of EFP (\ref{gauss}). However, for long enough $R$ it is the $1/T$ scale that defines the temporal size of the instanton, the space-time area of the disturbance scales as $R$ and one obtains
\be
	P(R) \sim e^{-\gamma R}.
 \la{exp}
\ee
One expects a crossover between the Gaussian and the exponential behavior taking place at $R\sim v_{s}/T$. The result (\ref{exp}) is very familiar from statistical physics with $\gamma$ measuring the difference (per unit length) of free energies between the true ground state of the liquid and the empty state. The Gaussian decay (\ref{gauss}) is a manifestation of an effective increase of the dimensionality from one to two in quantum systems at $T=0$.

In the above argument we used the assumption of compressibility of the quantum liquid. For incompressible liquids one has a finite correlation length of density fluctuations. The argument can be extended then to obtain an exponential form (\ref{exp}) with the correlation length playing the role of an inverse temperature \cite{abanovfran, franabanov}.


\section{Hydrodynamic approach}
\label{sec_hydro}

The collective description we are looking for is nothing else but a hydrodynamic description in terms of density $\rho$ and current $j=\rho v$ (or velocity $v$) of a one-dimensional liquid. For simplicity, let us consider first the case of a system with Galilean invariance. Then we write the Euclidian action of a liquid as
\be
	S = \int d^{2}x\, \left[\frac{j^{2}}{2\rho} +\rho\epsilon(\rho)+\ldots \right],
 \la{Shyd}
\ee
where the first term is fixed by the Galilean invariance and is the kinetic energy of the liquid moving as a whole. The second term is the internal energy of the fluid which is determined by the equation of state of the liquid. $\epsilon(\rho)$ is the internal energy per particle at given density $\rho$.  The terms denoted by dots are the terms which depend on density and its spatial derivatives. These terms will be small in the problems, where density and velocity gradients are small compared to $\rho_{0}$. Let us now remember that due to the particle conservation the density and current are not independent variables but are related by a constraint -- the continuity equation
\be
	\partial_{t}\rho+\partial_{x}j=0.
 \la{continuity}
\ee
One can easily solve the constraint (\ref{continuity}) introducing a particle displacement field $u$ such that
\bea
	\rho &=& \rho_{0}+\partial_{x}u,
 \nonumber \\
 	j &=& -\partial_{t}u.
 \la{displacement}
\eea
Microscopic definition of the displacement field is $u(x)+\rho_{0}x=\sum_{j}\theta(x-x_{j})$, where $\theta(x)$ is a step function and $x_{j}$-s are coordinates of particles.
It is easy to check that the configuration $u(x,t)$ minimizing $S[u]$ from (\ref{Shyd}) is given by $\delta S=0$ or after simple algebra
\be
 \la{euler}
	\partial_{t}v+v\partial_{x}v = \partial_{x} \partial_{\rho}[\rho\epsilon(\rho)],
\ee
which is the Euler equation of a one-dimensional hydrodynamics. The sign of the r.h.s. of (\ref{euler}) differs from the conventional minus sign because we work in the Euclidian formulation.

The action (\ref{Shyd}) with the parameterization (\ref{displacement}) provide us with a variational formulation of one-dimensional classical hydrodynamics. To calculate the probability of hydrodynamic fluctuations at zero temperature we have to quantize this hydrodynamics. To the best of my knowledge, the first ``quantization'' of hydrodynamics was done by L. D. Landau in Ref.\cite{Landau-1941}, where he used essentially (\ref{Shyd}) as a quantum Hamiltonian of the liquid with density and velocity fields satisfying commutation relations $\left[\rho(x),v(y)\right]=-i\partial_{x}\delta(x-y)$.
For the purpose of evaluating a rare fluctuation the path integral approach is more useful and we use the partition function (\ref{Z}), where $u(x,t)$ is the displacement field and the action $S[u]$ is given by (\ref{Shyd}) and (\ref{displacement}). We notice here that we did not specify the measure of integration $Du$ in (\ref{Z}). Finding this measure requires a derivation of an effective hydrodynamic formulation from the underlying microscopic physics. However, the ``non-flatness'' of the measure gives only gradient corrections similar to the already omitted terms denoted by dots in (\ref{Shyd}). Those corrections to the measure and the action will produce subdominant contributions to the asymptotics of EFP and will be neglected in this work.

Let us now summarize our strategy for the calculation of the leading term of the EFP. We  solve \textit{classical} equations of motion (\ref{continuity},\ref{euler}) with EFP boundary conditions (\ref{bcefp},\ref{bcinf}) and then find the value $S_{opt}$ of (\ref{Shyd}) on the obtained solution. Finally, (\ref{prinst}) will give us the dominant contribution to the EFP at $R\to\infty$.

We conclude this section with two remarks. First, it will be convenient for us to generalize the problem and replace the EFP boundary condition (\ref{bcefp}) by a slightly more general depletion formation probability (DFP) boundary condition \cite{abanovkor}
\be
	\rho(t=0;-R<x<R) =\bar\rho,
 \la{bcdfp}
\ee
where $\bar\rho$ is some constant density. In the case $\bar\rho=0$ we obtain the EFP problem while for $\bar\rho$ close to $\rho_{0}$ one can use the bosonization technique to calculate $P(R;\bar\rho)$.

Second remark is that one can easily obtain the functional dependence of $P(R)$ even without solving hydrodynamic equations (\ref{continuity},\ref{euler}) using very simple scaling arguments. Indeed, the equations (\ref{continuity},\ref{euler}) are uniform in space and time derivatives so that if $\rho(x,t)$ and $v(x,t)$ are solutions, then $\rho(\lambda x,\lambda t)$, $v(\lambda x,\lambda t)$ are also solutions of hydrodynamic equations. Choosing $\lambda =R$ we obtain the boundary condition (\ref{bcdfp}) as $\rho(t=0;-1<x<1) =\bar\rho$ and the only dependence on $R$ is left over in the integration measure of (\ref{Shyd}), which gives $S_{opt}\sim R^{2}$. Thus, we obtain a Gaussian decay of DFP (or EFP in particular) as a function of $R$. Corrections to this behavior come from the terms of higher order in gradients in the hydrodynamic action (denoted by dots in (\ref{Shyd})) as well as from fluctuations around the saddle point (classical) trajectory in the partition function (\ref{Z}).

\section{Linearized hydrodynamics or bosonization}
\label{sec_bosonization}

Before proceeding to the general case, let us consider the DFP for  
\be
 \la{smbos}
	\frac{\rho_{0}-\bar\rho}{\rho_{0}}\ll 1,
\ee
i.e., the probability of formation of a small constant density depletion along the long string $-R<x<R$. In this case the deviation of the density from $\rho_{0}$ is small almost everywhere (see below) and one can use a linearized version of generally nonlinear hydrodynamic equations (\ref{continuity},\ref{euler}). Expanding  the classical action (\ref{Shyd}) in gradients of the displacement field $u$ (\ref{displacement}) we obtain in harmonic approximation\footnote{The harmonic approximation to the nonlinear hydrodynamics of quantum liquid is equivalent to a linear bosonization approach to interacting one-dimensional particles.}
\be
 \la{Sbos}
	S_{bos} = \frac{v_{s0}}{\rho_{0}}\int d^{2}x\, \frac{1}{2}(\partial_{\mu}u)^{2},
\ee
where we scaled $v_{s0}t\to t$. The sound velocity $v_{s0}$ at $\rho=\rho_{0}$ is defined as
\be
	v_{s0}^{2} = \rho \partial_{\rho}^{2}(\rho\epsilon(\rho))\Big|_{\rho=\rho_{0}}.
\ee
The corresponding equation of motion is the Laplace equation
\be
 \la{laplace}
	\Delta u =0
\ee
with the DFP boundary condition
\be
 \la{bcdfpbos}
	u(x,t=0) = -(\rho_{0}-\bar\rho) x, \;\;\;\; \mbox{for}\; -R<x<R.
\ee

It is easy to see that the solution of (\ref{laplace},\ref{bcdfpbos}) decaying sufficiently fast at infinity is given by 
\be
 \la{bossol}
	u(x,t) = -(\rho_{0}-\bar\rho)\, \R\left(z_{0}-\sqrt{z_{0}^{2}-R^{2}}\right),
\ee 
where we introduced the complex notation $z_{0} = x+iv_{s0}t$ (where $t$ is the original imaginary time). Indeed, at $t=0$, $-R<x<R$ the complex coordinate $z_{0}$ is real and square root in (\ref{bossol}) is purely imaginary so that (\ref{bcdfpbos}) is satisfied.

At space-time infinity $z_{0}\to \infty$ we have 
\be
 \la{asymptu}
	u(x,t) \approx -\frac{\alpha}{z_{0}}-\frac{\bar\alpha}{\bar z_{0}}
\ee
with 
\be
 \la{alphabos}
	\alpha =\bar\alpha = \frac{1}{4}(\rho_{0}-\bar\rho)R^{2}.
\ee
We obtain from (\ref{asymptu}) that at $z_{0}\to\infty$
\bea
	\rho &\approx& \rho_{0} +\frac{\alpha}{z_{0}^{2}} +\frac{\bar\alpha}{\bar z_{0}^{2}},
 \nonumber \\
 	v &\approx& -i \frac{v_{s0}}{\rho_{0}} \left(\frac{\alpha}{z_{0}^{2}}
	-\frac{\bar\alpha}{\bar z_{0}^{2}} \right),
 \la{rvbosinf}
\eea
which obviously satisfy the boundary conditions at infinity (\ref{bcinf}).

Now that we obtained the solution of hydrodynamic equations it is a straightforward problem to calculate the value of the action (\ref{Sbos}) on this solution\footnote{See the next section on how to avoid doing this calculation.} and obtain
\be
 \la{dfpbos}
	S_{DFP} = \frac{1}{2}\frac{v_{s0}}{\pi\rho_{0}}\left[\pi(\rho_{0}-\bar\rho)R\right]^{2}.
\ee

We note that the linearization of hydrodynamic action (\ref{Sbos}) is not self-consistent near the ends of the string $t=0$, $x=\pm R$ where the solution (\ref{bossol}) is singular and gradients of $u$ diverge. However, one can estimate the corrections coming from those areas and find that they change the coefficient $[\pi(\rho_{0}-\bar\rho)]$ in front of the $R^{2}$ in (\ref{dfpbos}) by terms of the higher order in small parameter (\ref{smbos}) \cite{abanovkor}.

We also note here that although the value of the coefficient $\alpha$ given by (\ref{alphabos}) is approximate and is valid only in the limit of a very weak depletion, the asymptotic forms (\ref{asymptu},\ref{rvbosinf}) are very general as they depend only on the linearization of hydrodynamic equations at $x,t\to \infty$, where it is always possible.

\section{EFP through an asymptotics of the solution}
\label{sec_efp}

The calculation of the EFP has already been reduced to the calculation of the value of the classical action on the solution of equations of motion satisfying EFP boundary conditions. In this section we are going use a Maupertui principle \cite{LL-mech} to obtain a simple expression for $S_{opt}$ in terms of the asymptotics of the EFP solution of hydrodynamic equations.

Let us calculate the variation of the action (\ref{Shyd}) with respect to the displacement field $u$
\bea
	\delta S  &=& \int d^{2}x\, \Big\{-\partial_{t}(v\delta u) 
	-\partial_{x}\left[\left(\frac{v^{2}}{2}-\partial_{\rho}(\epsilon\rho)\right)
	\delta u \right]
 \nonumber \\
	&+& \delta u\left[\partial_{t}v +v\partial_{x}v 
	-\partial_{x}\partial_{\rho}(\epsilon\rho)\right] \Big\}.
 \la{dS1}
\eea
We kept here surface terms (full derivatives) in addition to the last term which produces the equation of motion (\ref{euler}). Now we assume that the action $S$ is calculated on the EFP (or DFP) solution of equations of motion and consider the derivative of this action with respect to the equilibrium background density $\rho_{0}$. We have
\bea
	\partial_{\rho_{0}}S_{opt}
	=\int d^{2}x\, \left\{-\partial_{t}(v\partial_{\rho_{0}} u) 
	-\partial_{x}\left[\left(\frac{v^{2}}{2}-\partial_{\rho}(\epsilon\rho)\right)
	\partial_{\rho_{0}}u\right]\right\},
 \la{dS2}
\eea
where we used the fact that $u$ satisfies the Euler equation and dropped the last term in (\ref{dS1}). (\ref{dS2}) has only derivative terms and can be re-written as a boundary contribution
\bea
	\partial_{\rho_{0}}S_{opt}
	=\oint \, \left[v\partial_{\rho_{0}} u\, dx 
	+\left(\partial_{\rho}(\epsilon\rho)-\frac{v^{2}}{2}\right)
	\partial_{\rho_{0}}u \,dt \right],
 \la{dS3}
\eea
where the integral is taken over the infinitely large contour around $xt$ plane.
As the integrand in (\ref{dS3}) should be calculated at (infinitely) large $x$ and $t$ we can use in (\ref{dS3}) the general asymptotics (\ref{asymptu},\ref{rvbosinf}). After simple manipulations we obtain our main result
\be
 \la{main}
	\partial_{\rho_{0}}S_{opt} = 2\pi \frac{v_{s0}}{\rho_{0}}(\alpha+\bar{\alpha}).
\ee

As a simple check of this result we take the value of $\alpha$ obtained in bosonization approach (\ref{alphabos}) and substitute it into (\ref{main}). We immediately obtain $\partial_{\rho_{0}}S_{opt} = \pi \frac{v_{s0}}{\rho_{0}} (\rho_{0}-\bar\rho)R^{2}$,  which is equivalent to (\ref{dfpbos}) up to the terms of higher order in the  small parameter (\ref{smbos}).

\section{Free fermions}
\label{sec_free}

In this section we find the EFP for a free Fermi gas in one dimension. First, let us find the internal energy $\epsilon(\rho)$ of the gas. The density of fermions is given in terms of Fermi momentum $k_{F}$ as
\be
	\rho =\int _{-k_{F}}^{k_{F}}\frac{dk}{2\pi} = \frac{k_{F}}{\pi}.
\ee
The energy per unit length is
\be
	\rho\epsilon(\rho) =\int _{-k_{F}}^{k_{F}}\frac{k^{2}}{2}\,\frac{dk}{2\pi} 
	= \frac{k_{F}^{3}}{6\pi} =\frac{\pi^{2}}{6}\rho^{3}
\ee
where we put $\hbar=1$ and fermion mass $m=1$. The energy per particle in a free Fermi gas with density $\rho$ is 
\be
	\epsilon(\rho) = \frac{\pi^{2}}{6}\rho^{2}.
 \la{erff}
\ee
We calculate the sound velocity using 
\be
	v_{s}^{2} = \rho \partial_{\rho}^{2}(\rho\epsilon)
\ee
and obtain
\be
 \la{vsff}
	v_{s} = \pi\rho =k_{F}
\ee
-- the well known result that the sound velocity and Fermi velocity are the same (remember that in our notations $m=1$ and $v_{F}=k_{F}$).

\begin{figure}[t]
 \begin{center}
   \includegraphics[scale=0.25]{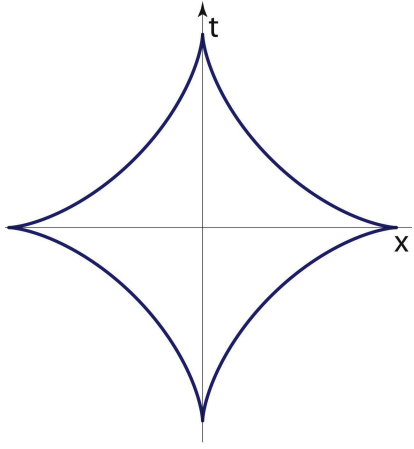}
 \end{center}
   \caption{The region of the $x-t$ plane in which $\rho(x,t)=0$ for the EFP instanton for free fermions is shown. The boundary of the region is given by an astroid $x^{2/3}+(\pi\rho_{0}t)^{2/3}=R^{2/3}$.}
 \label{fig:droplet} 
\end{figure}

Hydrodynamic equations (\ref{continuity}) and (\ref{euler}) for free fermions are
\bea
	\partial_{t}\rho +\partial_{x}(\rho v) &=& 0,
 \nonumber \\
 	\partial_{t}v + v\partial_{x}v &=& \pi^{2}\rho\partial_{x}\rho.
 \la{hff}
\eea
Introducing a complex field 
\be
 \la{w}
	w = \pi \rho +iv,
\ee
we re-write both equations (\ref{hff}) as a single complex Hopf equation\footnote{In real time formalism, instead of $w,\bar{w}$ one introduces ``right and left Fermi momenta'' $k_{R,L}=\pi\rho\pm v$ which satisfy the Euler-Hopf equations $\partial_{t}k+k\partial_{x}k=0$ reflecting the absence of interactions between fermions.}
\be
 \la{hopf}
 	\partial_{t}w -iw\partial_{x}w =0.
\ee
The latter equation has a general solution
\be
 \la{hsol} 
	w = F(z),
\ee
where $F(z)$ is an arbitrary analytic function of a complex variable $z$ defined as 
\be
 \la{zhopf}
 	z = x+iwt.
\ee
The boundary conditions will determine a particular analytic function $F(z)$ so that the equations (\ref{hsol},\ref{zhopf}) will define the solution $w(x,t)$ of (\ref{hopf}) implicitly. It is easy to check that the unknown function $F(z)$ for DFP is given by
\be
 \la{Fff}
	F(z) = \pi\bar\rho +\pi(\rho_{0}-\bar\rho) \frac{z}{\sqrt{z^{2}-R^{2}}}
\ee
and the one for EFP can be obtained by putting $\bar\rho=0$ in (\ref{Fff}). The space-time configuration of the density and velocity fields minimizing the hydrodynamic action is given by real and imaginary parts of $w(x,t)$ (see Eq. (\ref{w})) which is implicitly defined by
\be
 \la{wffsol}
	w -  \pi\bar\rho = +\pi(\rho_{0}-\bar\rho) \frac{z}{\sqrt{z^{2}-R^{2}}}
\ee
with (\ref{zhopf}). 
\begin{figure}[t]
 \begin{center}
   \includegraphics[scale=0.5]{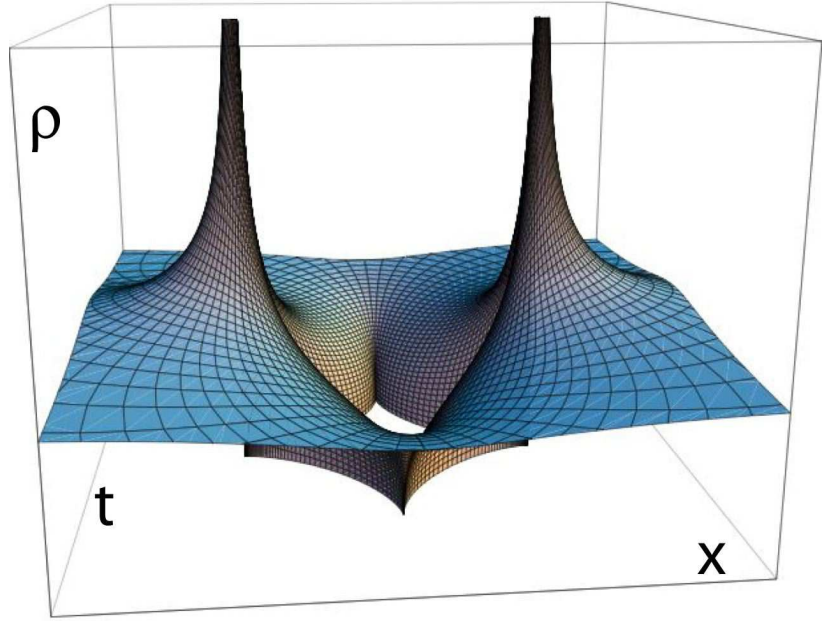}
 \end{center}
   \caption{The density profile $\rho(x,t)$ is shown for the EFP instanton. The density diverges at points $(x,t)=(\pm R,0)$. The shape of the ``Emptiness'' is shown in Fig.\ref{fig:droplet}.}
 \label{fig:density} 
\end{figure}
This solution is relatively complicated (see Figs. \ref{fig:density}, \ref{fig:droplet}) and a direct calculation of the value of the hydrodynamic action (\ref{Shyd}) on this solution is cumbersome. However, one can get the result very quickly using (\ref{main}). 
Indeed, in the limit $x,t\to \infty$ we have $w\to \pi\rho_{0}$ and $z=x +iwt\to x+i\pi\rho_{0}t =z_{0}$. Therefore, the asymptotics of  (\ref{wffsol}) is given by
\be
 \la{wasff}
 	w-\pi\rho_{0} \approx \pi(\rho_{0}-\bar\rho) \frac{R^{2}}{2z_{0}^{2}}
\ee
and taking, e.g., its real part
\be
 \la{rhoffas}
	\rho -\rho_{0} \approx \frac{1}{4}(\rho_{0}-\bar\rho) R^{2}
	\left(\frac{1}{z_{0}^{2}}+\frac{1}{\bar{z}_{0}^{2}}\right).
\ee
Comparing (\ref{rhoffas}) with (\ref{rvbosinf}) we extract
\be
	\alpha = \frac{1}{4}(\rho_{0}-\bar\rho) R^{2}
\ee
and substituting into (\ref{main}) we derive $\partial_{\rho_{0}}S_{opt}=\pi^{2}(\rho_{0}-\bar\rho)R^{2}$ and 
\be
 \la{soptff}
 	S_{opt} = \frac{1}{2}\left[\pi(\rho_{0}-\bar\rho)R\right]^{2}.
\ee
This gives for DFP and EFP probabilities respectively
\bea
	P_{DFP}(R) &\sim& \exp\left\{
	-\frac{1}{2}\left[\pi(\rho_{0}-\bar\rho)R\right]^{2} \right\},
 \la{dfpff} \\
	P_{EFP}(R) &\sim& \exp\left\{
	-\frac{1}{2}\left(\pi \rho_{0}R\right)^{2} \right\}.
 \la{efpff} 
\eea

In the case of free fermions an exact asymptotic expansion of EFP in $1/R$ is known \cite{dysmehta} with first few terms given by (\ref{FreeFermionsExact}). The instanton contribution (\ref{efpff}) gives an exact first (Gaussian) term of this expansion.

\section{Calogero-Sutherland model}
\label{sec_calogero}

Our next example of a one-dimensional liquid is the Calogero-Sutherland model -- a model of one-dimensional particles interacting with an inverse square potential. The Hamiltonian of the model is
\bea
	H &=& -\frac{1}{2}\sum_{j=1}^{N}\frac{\partial^{2}}{\partial x_{j}^{2}}
	+\frac{1}{2} \sum_{1\leq j<k\leq N} \frac{\lambda(\lambda-1)}{(x_{j}-x_{k})^{2}}
 \la{CSh} \\
	&=& -\frac{1}{2}\sum_{i=1}^{N} \left(\frac{\partial}{\partial x_{i}}
	+\sum_{j=1, j\neq i}^{N}\frac{\lambda}{x_{i}-x_{j}} \right)
	\left(\frac{\partial}{\partial x_{i}}
	-\sum_{k=1, k\neq i}^{N}\frac{\lambda}{x_{i}-x_{k}} \right).
 \nonumber
\eea
Here we again use the units $\hbar=1$ and $m=1$. This model is known to be integrable \cite{Calogero-1969,Sutherland-1971}. We are interested in the thermodynamic limit of (\ref{CSh}). The easiest way to go to this limit is to consider (\ref{CSh}) in an additional harmonic potential which does not destroy the integrability of the model and then take a limit of number of particles $N\to \infty$ and the strength of the potential going to zero so that the density is kept constant and equal $\rho_{0}$. We omit all these details which can be found in the original papers \cite{Calogero-1969,Sutherland-1971}. We mention only that the ground state wave function of (\ref{CSh}) is 
\be
	\Psi_{GS} = \prod_{j<k}(x_{j}-x_{k})^{\lambda}
\ee
and shows an intermediate statistics interpolating between non-interacting bosons ($\lambda=0$) and non-interacting fermions ($\lambda=1$). We also notice here that the probability distribution of particle coordinates
\be
	\left|\Psi_{GS}\right|^{2} = \prod_{j<k} |x_{j}-x_{k}|^{2\lambda}
\ee
at particular values of coupling constant $\lambda =1/2,\, 1,\, 2$ coincides with the 
joint probability of eigenvalues for the orthogonal, unitary, and symplectic random matrix ensembles respectively (see Eq. (\ref{jed})).

To calculate the leading behavior of the EFP for the Calogero-Sutherland model we need the internal energy $\epsilon(\rho)$ which can be easily found \cite{Sutherland-1971}
\be
	\epsilon(\rho) = \frac{\pi^{2}}{6}\lambda^{2} \rho^{2}.
 \la{ercs}
\ee
The  (\ref{ercs}) differs from the free fermion case (\ref{erff}) by a factor of $\lambda^{2}$ and coincides with the latter (as expected) at $\lambda=1$. Introducing 
\be
	w=\lambda\pi\rho +iv
\ee
and repeating the calculations of the previous section we obtain
\bea
	P_{DFP}(R) &\sim& \exp\left\{
	-\frac{1}{2}\lambda \left[\pi(\rho_{0}-\bar\rho)R\right]^{2} \right\},
 \la{dfpcs} \\
	P_{EFP}(R) &\sim& \exp\left\{
	-\frac{1}{2}\lambda \left(\pi\rho_{0}R\right)^{2} \right\}.
 \la{efpcs} 
\eea
Comparing  to the known exact result (\ref{CalogeroExact}) we see that (\ref{efpcs}), indeed, gives the exact leading asymptotics of the EFP for the Calogero-Sutherland model. Subleading (in $1/R$) corrections to (\ref{efpcs}) are due to gradient corrections to the hydrodynamic action (\ref{ercs}) and to quantum fluctuations around the found instanton.

\section{Free fermions on the lattice}
\label{sec_lattice}

The goal of this section is to illustrate that the hydrodynamic method we used is not limited to Galilean invariant systems. We use the method to calculate the EFP for a system of non-interacting lattice fermions with an arbitrary dispersion $\varepsilon(k)$. 
Semiclassically, one can describe the evolution of the degenerate 1D Fermi gas in terms of two smooth and slow functions $k_{R,L}(x,t)$ -- right and left Fermi points respectively. The equations of motion are given by
\be
	\partial_{t}k_{R,L} +\varepsilon'(k_{R,L}) \partial_{x}k_{R,L} =0,
 \la{emlff}
\ee
which is an obvious consequence of the absence of interactions. Indeed, the derivative $\varepsilon'(k_{R,L})$ is the right (or left) Fermi velocity of particles, and  (\ref{emlff}) is nothing else but the statement that the momentum of a particle does not change in time. The classical action of 1D Fermi gas should reproduce (\ref{emlff}) as well as give the correct energy of the system $E=\int dx\,\int_{k_{L}(x)}^{k_{R}(x)}\frac{dk}{2\pi}\,\varepsilon(k)$.

We assume here that the dispersion of fermions is parity invariant $\epsilon(k)=\epsilon(-k)$ and consider the following Euclidian action
\be
 \la{aw}
    S = \int d^{2}x\, \left\{ \frac{i}{4\pi}\left[
    w\partial_{x}^{-1}\partial_{t}w 
    -\bar w\partial_{x}^{-1}\partial_{t}\bar w \right]
    +\int_{-\bar w}^{w}\frac{dk}{2\pi}\, \varepsilon(k) \right\}.
\ee
Here $\varepsilon(k)$ is the dispersion (including chemical potential) of free fermions and $w,\bar w$ are independent variables which are Euclidian versions of $k_{R}$, $-k_{L}$. We identify the density of particles as $\rho = \frac{w+\bar w}{2\pi}$. The equations of motion corresponding to (\ref{aw})
\bea
    i\partial_{t}w+\partial_{x}\varepsilon(w) &=& 0,
 \nonumber \\
    -i\partial_{t}\bar w+\partial_{x}\varepsilon(\bar w) &=& 0
 \la{fmeq}   
\eea 
are, indeed, the Euclidian versions of (\ref{emlff}). Equations (\ref{fmeq}) are complex conjugates of each other ($\overline{\varepsilon(w)}=\varepsilon(\bar{w})$ as $\varepsilon(k)$ is a real function). The difference of these two equations give the continuity equation (\ref{continuity}) with the current $j=\frac{\varepsilon-\bar\varepsilon}{2\pi i}$. The path integral with the action (\ref{aw}) is the integral over two fields $w,\bar{w}$. We integrate out their difference $w-\bar{w}$ in a saddle point approximation. Namely, we re-write (\ref{aw}) in terms of combinations $w-\bar{w}$ and $w+\bar{w}$. Then, we take a variation of the action with respect to $w-\bar{w}$. This produces the continuity equation relating $w$ and $\bar{w}$. We solve the continuity equation introducing the displacement field.  Thus, we have $w$ and $\bar{w}$ in terms of $u$ as the saddle point trajectory. Namely, 
\bea
    \rho &=& \frac{w+\bar w}{2\pi} = \rho_{0} +\partial_{x}u,
 \nonumber \\
    j &=& \frac{\varepsilon-\bar\varepsilon}{2\pi i} = -\partial_{t}u.
 \la{rjulf}
\eea
We substitute these expressions back into (\ref{aw}) and obtain
\be
 \la{awint}
    S[u] = \int d^{2}x\, \left\{ -\frac{1}{4\pi}(w-\bar{w})
    \left[\varepsilon(w)-\varepsilon(-\bar{w})\right]
    +\int_{-\bar w}^{w}\frac{dk}{2\pi}\, \varepsilon(k) \right\}.
\ee
We assume now that the integration variable of the path integral is the displacement field $u$. The fields $w,\bar{w}$ in (\ref{awint}) are not independent but are related to $u$ by (\ref{rjulf}). Deriving (\ref{awint}) we neglected the fluctuations around the saddle point as well as the changes in the measure of path integration. The corrections due to the neglected terms will be of the higher order in field gradients and are not essential for our calculation of the leading term of the EFP.

One can now use (\ref{awint}) in the path integral formulation of the quantum hydrodynamics (\ref{Z}) where the integration is taken over all configurations of the displacement field $u$. We note here that in the Galilean invariant systems the ``kinetic term'' of the hydrodynamic action (\ref{Shyd}) is fixed. The Lagrangian of the action (\ref{awint}) is a more complicated function of $\partial_{t}u$. In this case we derived it using the fact that fermions are free. In a more general problem of interacting particles the derivation of the hydrodynamic action requires the solution of the dynamic many body problem. 

The Lagrangian of (\ref{awint}) is given by
\be
    L =-\frac{1}{4\pi}(w-\bar{w})
    \left[\varepsilon(w)-\varepsilon(-\bar{w})\right]
    +\int_{-\bar w}^{w}\frac{dk}{2\pi}\, \varepsilon(k).
 \la{lagrlf}
\ee
Now, we calculate the EFP for lattice fermions using the the Lagrangian (\ref{lagrlf}) with the Eq. (\ref{rjulf}), and the results of Appendix A. First, we calculate
\be
    dL =\frac{\varepsilon+\bar\varepsilon}{2}d\rho
    +\frac{w-\bar w}{2i}dj
\ee
and
\bea
    L_{j} &=& v = \frac{w-\bar w}{2i},
 \la{vlf} \\
    L_{\rho} &=& \frac{\varepsilon+\bar\varepsilon}{2},
 \\
    L_{jj} &=& \frac{2\pi}{\varepsilon'+\bar\varepsilon'},
 \\
    L_{\rho j} &=& -\frac{2\pi}{\varepsilon'+\bar\varepsilon'}
    \frac{\varepsilon'-\bar\varepsilon'}{2i},
 \\
    L_{\rho\rho} &=& \frac{2\pi}{\varepsilon'+\bar\varepsilon'}|\varepsilon'|^{2},
 \la{lrrlf}
\eea
where $\varepsilon'= \left|\frac{\partial\varepsilon}{\partial k}\right|_{k=w}$ and $\bar\varepsilon'= \left|\frac{\partial\varepsilon}{\partial k}\right|_{k=\bar w}$. In terms of the density and velocity
\be
	w=\pi \rho +iv.
 \la{wrvlf}
\ee
We emphasize that because of the absence of the Galilean invariance (the dispersion is  {\it not} $k^{2}/2m$), the current is {\it not} $\rho v$ but can be found from (\ref{rjulf},\ref{vlf}).

We obtain then an interesting relation
\be
     \kappa \equiv \sqrt{L_{\rho\rho}L_{jj}-L_{\rho j}^{2}} 
     =\pi.
 \la{piid}
\ee
This relation can be considered as a special property of free fermions (with arbitrary dispersion!). Namely, one can trace the origin of (\ref{piid}) to the fact that each fermion occupies a fixed volume in the phase space. \footnote{For Calogero-Sutherland model (\ref{ercs}) we have $\kappa =\pi\lambda$. It means that the volume of the phase space per particle is changed by the factor of $\lambda$ which reflects the fractional statistics of particles.}

Let us now apply Riemann's trick to equations (\ref{fmeq}). Namely, we interchange independent and dependent variables so that $x=x(w,\bar w)$ and $t=t(w,\bar w)$. Then using (\ref{piid}, \ref{vlf}-\ref{lrrlf}) we re-write (\ref{fmeq}) as
\be
   \partial_{w}(x-i\bar\varepsilon' t) =0
 \la{fmeqR}
\ee
and its complex conjugate. The most general solution of (\ref{fmeqR}) is given implicitly by
\be
    w = F(z),
 \la{wfz}
\ee
where $F(z)$ is an arbitrary analytic function and 
\be
    z = x +i\varepsilon'(w) t.
 \la{zw}
\ee
The equations (\ref{fmeqR}, \ref{zw}) reduce to equations for free continuous fermions if $\varepsilon(k)=k^{2}/2$.

Similarly to free continuous fermions, the EFP boundary conditions specify the form of the function $F(z)$ in (\ref{wfz}).
We define
\bea
    \tilde{w} &=& \frac{w-\pi\bar\rho}{2(1-\bar\rho)},
 \\
    \tilde{\theta} &=& \frac{\pi(\rho_{0}-\bar\rho)}{2(1-\bar\rho)},
\eea
where we assume lattice spacing to be 1 so that the maximal density of fermions on the lattice is $1$.
Then the EFP solution is given by $F(w)$ such that
\be
     \sin\tilde{w} = \sin\tilde{\theta}
     \frac{z}{\sqrt{z^{2}-R^{2}}}.
\ee
We immediately find at large distances ($z\to z_{0}=x+i\varepsilon_{0}' t \to \infty$)
\be
     w-\pi\rho_{0} \approx 2(1-\bar\rho)\frac{R^{2}}{2z_{0}^{2}}\tan\tilde{\theta}
\ee
and using (\ref{rvbosinf}) and (\ref{wrvlf}) we extract
\be
    \alpha = \frac{1}{2\pi}(1-\bar\rho)R^{2}\tan\tilde\theta.
\ee
We substitute this expression in (\ref{maingen}) and obtain
\be
    \partial_{\rho_{0}}S =2\pi(1-\bar\rho)R^{2}
    \tan\frac{\pi(\rho_{0}-\bar\rho)}{2(1-\bar\rho)},
\ee
so that
\be
 \la{dfpLat}
    S_{DFP} = -4(1-\bar\rho)^{2}R^{2}
    \ln\cos\frac{\pi(\rho_{0}-\bar\rho)}{2(1-\bar\rho)}.
\ee
In particular for $\bar\rho=0$ we obtain for the Emptiness Formation Probability
\be
 \la{efpLat}
    S_{EFP} = -4R^{2}
    \ln\cos\frac{\pi\rho_{0}}{2}
    = -4R^{2}
    \ln\cos\frac{k_{F}}{2}.
\ee
This result is the exact first term in $1/R$ expansion (\ref{exlf}). The next term is $\frac{1}{4}\ln R$.
It is interesting to note that as it should be, in the limit $\rho_{0}-\bar\rho\ll \rho_{0}$ the result (\ref{dfpLat}) reproduces the bosonization result (\ref{dfpbos}). Also, in the limit $\rho_{0},\bar\rho\ll 1$ the results (\ref{dfpLat},\ref{efpLat}) reproduce the corresponding results (\ref{dfpff},\ref{efpff}) for free continuous fermions.

\section{Conclusion}

We showed on the example of the Emptiness Formation Probability that one can obtain some exact results for correlation functions using a collective hydrodynamic description. Moreover, the nonlinear hydrodynamic description might work and produce non-perturbative results even in cases where the linearized hydrodynamics fails. We obtained the leading term of the asymptotics of the EFP using the instanton approach. Although the EFP is a property of the ground state of a quantum many body system, the hydrodynamic approach can also be used to study dynamics of quantum systems. It is also not limited to integrable systems.

We considered few simple systems which are in the Luttinger liquid phase at zero temperature. The effects of finite temperature and finite gap in the spectrum of excitations can also be considered using the hydrodynamic approach in the limit when temperature is very low and the gap is very small \cite{abanovkor,franabanov}. In these limits one obtains a crossover between the Gaussian decay of the EFP at intermediate $R$ to the exponential decay at very large $R$.

There are also many questions which are left open. First, it would be nice to obtain the results for the EFP in other integrable systems such as bosons with delta-repulsion and XXZ spin chains. Some results for the EFP have already been obtained using other methods (see Appendix B), while, e.g., the EFP for a XXZ spin chain in the presence of magnetic field is still not known. The application of the hydrodynamic approach developed here to these systems is straightforward and reduces the problem to the problem of finding asymptotics of the solution of a system of classical equations with proper boundary conditions. However, these classical equations are complicated and we haven't yet obtained analytical results for their asymptotics.

The second open question involves the corrections to the leading hydrodynamic approximation that we used in this lecture. There are two sources of such corrections: the gradient corrections to the ``classical'' hydrodynamic action and quantum fluctuations around the classical saddle point. E.g., if one is interested in the next to leading terms of the asymptotic expansion of the EFP in $1/R$ one needs to include these gradient corrections (which make hydrodynamics dispersive) and quantum fluctuations. Especially interesting would be to obtain the power law pre-factor of the EFP (or $\ln R$ term in the asymptotic expansion). The hydrodynamic calculation might shed some light on a possible universality of the exponent of this pre-factor.

Needless to say, that although we focused in this lecture on the particular correlation function (the EFP), the use of hydrodynamic approach is much broader. In particular, the nonlinear hydrodynamics is important to capture a lot of important nonlinear phenomena which disappear in the linear approximation.

\begin{acknowledgments}

I would like to thank the organizers of the 2004 Les Houches Summer School on Applications of Random Matrices in Physics for a very nice atmosphere of the School.
I have greatly benefited from discussions with F. Essler, F. Franchini,
V. E. Korepin, S. L. Lukyanov, A. M. Tsvelik, and P. B. Wiegmann.
This work was supported by the NSF grant DMR-0348358, and the
Theory Institute of Strongly Correlated and Complex Systems at
Brookhaven.

\end{acknowledgments}

\chapappendix{Hydrodynamic approach to non-Galilean invariant systems}
\label{app_nGal}

The hydrodynamic description of Sec. \ref{sec_hydro} and the approach of Sec. \ref{sec_efp} can be easily generalized to non-Galilean invariant systems. We briefly list here relevant formulas leaving the details of (straightforward) calculations to the interested reader. We assume that a one-dimensional compressible liquid can be described by the partition function (\ref{Z}) where the functional integration is taken over all configurations of the displacement field $u$ which defines the hydrodynamic density and current by (\ref{displacement}) and where the Euclidian action (\ref{act}) can be written as
\be
	S = \int d^{2}x\, L(\rho,j).
 \la{actgen}
\ee
The Lagrangian density $L(\rho,j)$ is a function of $\rho$ and $j$. As in the main text we neglect the gradient corrections to the action (\ref{actgen}). In the case of non-Galilean invariant systems these gradient corrections will depend on gradients of current in addition to gradients of density. The case of Galilean invariant systems considered in the main text (\ref{Shyd}) can be obtained from the general one assuming the particular Galilean invariant form  of the Lagrangian $L(\rho,j) =j^{2}/2\rho +\rho\epsilon(\rho)$. 

Variation of (\ref{actgen}) with respect to $u$ gives a generalized Euler equation which together with the continuity equation reads
\bea
	\partial_{t}\rho +\partial_{x}j &=& 0,
 \la{contgen} \\
	\partial_tL_j-\partial_x L_\rho  &=& 0,
 \la{eulergen}
\eea
where $L_{j}$ means $\partial L/\partial j$ etc.

If the deviation from the ground state is small, we can expand the action (\ref{actgen}) around $\rho=\rho_{0}$, $j=0$. We assume that $L_{\rho j}^{(0)}=0$, where superscript $(0)$ means that the derivative is calculated at the equilibrium values $\rho=\rho_{0}$ and $j=0$. We obtain for the action in harmonic approximation
\be
    S = \int d^{2}x\,\left[L_{jj}^{(0)}u_{t}^{2} +L_{\rho\rho}^{(0)}u_{x}^{2} \right].
\ee
We introduce a complex space-time coordinate
\bea
    z_{0} = x+iv_{s0}t,
 \la{z0gen}
\eea
where 
\be
 \la{vs0}
    v_{s0} = \sqrt{\frac{L_{\rho \rho}^{(0)}}{L_{jj}^{(0)}}}
\ee
is the sound velocity at equilibrium. Let us define the coefficient $\alpha$ through the asymptotics of an EFP solution at $z_{0}\to\infty$ by (\ref{asymptu}). The asymptotics of the current and density are given by (\ref{rvbosinf}).

Our main formula (\ref{main}) for the variation of the EFP instanton action with respect to the background density becomes
\be
 \la{maingen}
    \partial_{\rho_{0}}S = 2\pi \kappa_{0}\,
    (\alpha+\bar\alpha),
\ee  
where $\kappa_{0}$ is given by
\be
 \la{kappa0}
    \kappa_{0} =\sqrt{L_{\rho\rho}^{(0)}L_{jj}^{(0)}}.
\ee
The coefficient $\kappa_{0}$ is related to the compactification radius $R_{\rm comp}$ of bosons in the bosonization procedure  $\kappa_{0}=(2\pi R_{\rm comp})^{2}$. For free fermions $\kappa_{0}=\pi$ which corresponds to $R_{\rm comp}=1/\sqrt{4\pi}$.

\chapappendix{Exact results for EFP in some integrable models}
\label{app_exact}

For the sake of reader's convenience in this appendix we list some results obtained for the Emptiness Formation Probability $P(R)$ in integrable one-dimensional systems. 
We present the results for $S\equiv-\ln P(R)$ which should be compared with the instanton action $S_{opt}$ used in the main text.

\paragraph*{Free continuous fermions}

Let us denote 
\be
    s \equiv \pi\rho_{0}R.
 \la{s1}
\ee
We use the fact that the ground state wave function (more precisely $|\Psi|^{2}$) of free fermions coincides with the joint eigenvalue distribution of unitary random ensemble. For the latter the probability of having no eigenvalues in the range $2R$ of the spectrum was obtained in \cite{dysmehta} (see also \cite{mehta}). First few terms of the expansion in $1/R$ are
\be
	S = \frac{1}{2}s^{2} + \frac{1}{4}\ln s -\left(\frac{1}{12}\ln 2 +3\zeta'(-1)\right) 
	+O(s^{-2}).
 \la{FreeFermionsExact}
\ee

\paragraph*{Calogero-Sutherland model}

The Calogero-Sutherland model \cite{Calogero-1969, Sutherland-1971} (rational version, known also as Calogero model) with $N$-particles is defined as\footnote{To prevent particles running to infinity we either add  a harmonic potential to (\ref{1}) or put particles on a circle of a large radius.}
\be
 \label{1}
	{\cal H} = \frac{1}{2}\sum_{j=1}^{N}p_j^{2}
	+\frac{1}{2}\sum_{j,k=1; j\neq k}^{N}
	\frac{\lambda(\lambda-1)}{(x_{j}-x_{k})^{2}},
\ee
where $p_{j}=-i\partial/\partial x_{j}$ is the momentum operator of $j$-th particle and $\lambda$ is a dimensionless coupling constant. The wave function of the ground state is proportional to $\prod_{j<k}(x_{j}-x_{k})^{\lambda}$.  At $\lambda=1$ we have free fermions, while in the case of general $\lambda$ the model (\ref{1}) describes particles with fractional statistics. 
Using the form of the ground state wave function and thermodynamic arguments \cite{mehta} one obtains
\be
    S = \frac{\lambda}{2}s^{2} + (1-\lambda)s +O(\ln s).
 \la{CalogeroExact}
\ee
or defining 
\be
    s \equiv \sqrt{\lambda} \pi\rho_{0}R.
 \la{slambda}
\ee
and 
\be
	\alpha_{0} \equiv \frac{\lambda^{1/2}-\lambda^{-1/2}}{2}
 \la{alpha0}
\ee
we have 
\be
    S = \frac{1}{2}s^{2} - 2\alpha_{0} s +O(\ln s).
 \la{CalogeroExacta}
\ee
The notation $\alpha_{0}$ originates from the conformal field theory with central charge $c=1-24\alpha_{0}^{2}$ which is known to be related to the Calogero-Sutherland model \cite{Awata}. As far as I know the coefficient in front of $\ln s$ term of the expansion is not known for the general $\lambda$. However,  $\lambda =1/2, 1, 2$ correspond to random matrix ensembles where the full asymptotic expansion is known (see below). In those cases the coefficient of $\ln s$ is $1/8, 1/4, 1/8$ respectively. The natural guess is that 
\be
	S = \frac{1}{2}s^{2} - 2\alpha_{0} s 
	+\left(\frac{1}{4}-\alpha_{0}^{2}\right)\ln s +O(1).
 \la{CSguess}
\ee 

\subsubsection*{Random matrices}

For Random Matrix ensembles with $\beta=1,2,4$ the joint eigenvalue distribution is proportional to $\prod_{i<j}|z_{i}-z_{j}|^{\beta}$. The full asymptotic expansion of the quantity $E_{\beta}(0,2R)$ corresponding to the EFP $P(R)$ was obtained using properties of Toeplitz determinants \cite{dysmehta,mehta}. The first few terms of these expansions are given by
\bea
	S_{\lambda=1/2} &=& \frac{1}{4}(\pi\rho_{0}R)^{2} +\frac{1}{2}(\pi\rho_{0}R)
	+\frac{1}{8} \ln(\pi\rho_{0}R) -\frac{7}{24}\ln 2 - \frac{3}{2}\zeta'(-1) +O(1/s),
 \nonumber \\
	S_{\lambda=1} &=& \frac{1}{2}(\pi\rho_{0}R)^{2} 
	+\frac{1}{4} \ln(\pi\rho_{0}R) -\frac{1}{12}\ln 2 - 3\zeta'(-1) +O(1/s^{2}),
 \la{rmexact} \\
	S_{\lambda=2} &=& (\pi\rho_{0}R)^{2} -(\pi\rho_{0}R)
	+\frac{1}{8} \ln(\pi\rho_{0}R) +\frac{4}{3}\ln 2 - \frac{3}{2}\zeta'(-1) +O(1/s).
 \nonumber 
\eea 
Here we used $\lambda=\beta/2=1/2,1,2$ instead of $\beta$.
Using notations (\ref{slambda},\ref{alpha0}) we can summarize the first three terms of (\ref{rmexact}) in a compact form (\ref{CSguess}).

\subsubsection*{Free fermions on the lattice}

For non-interacting one-dimensional fermions on the lattice (and the corresponding XY spin chain) the asymptotic behavior of EFP was derived in \cite{shiroishi} using the Widom's theorem on the asymptotic behavior of Toeplitz determinants. Introducing the Fermi momentum $k_{F}=\pi\rho_{0}$ and using units in which the lattice spacing is $1$ we have  
\be
    S =-4R^{2}\ln\cos\frac{k_{F}}{2} + \frac{1}{4}\ln \left[2R\sin\frac{k_{F}}{2}\right]
    -\left(\frac{1}{12}\ln 2 +3\zeta'(-1)\right) +O(R^{-2}).
 \la{exlf}
\ee
In the continuous limit $k_{F}\to 0$ the (\ref{exlf}) goes to its continuous version (\ref{FreeFermionsExact}).

\subsubsection*{Bosons with delta repulsion}

The model of bosons with short range repulsion is described by
\be
 \label{2}
	{\cal H} = \frac{1}{2}\sum_{j=1}^{N}p_j^{2}
	+g \sum_{1\leq j< k\leq N}
	\delta(x_{j}-x_{k}),
\ee
where $g$ is a coupling constant.
It is integrable by Bethe Ansatz \cite{LiebLiniger-1963}. 
It was derived (conjectured) in Ref. \cite{ItsKorepinWaldron-1995} that the leading term of the EFP is
\be
    S = \frac{1}{2}(K R)^{2} \left[1+I(g/K)\right],
 \la{IKW}
\ee
where $K$ is the Fermi momentum in the Lieb-Liniger solution \cite{LiebLiniger-1963} and
\be
    I(x) = \frac{1}{2\pi^{2}} \int_{-1}^{1}\frac{y\,dy}{\sqrt{1-y^{2}}}\,
    \int_{-1}^{1}\frac{z\,dz}{\sqrt{1-z^{2}}}\,
    \log\left(\frac{x^{2}+(y+z)^{2}}{x^{2}+(y-z)^{2}}\right).
\ee
The limit $I(x\to \infty) = 0$ corresponds to the free fermion result (\ref{FreeFermionsExact}) (Tonks-Girardeau gas), while the limit $I(x\to 0) = 1$ is the result for the EFP of free bosons.

\subsubsection*{XXZ model}

The Hamiltonian of an XXZ model is given by
$$ H = J \sum_{k}\left[\sigma_{k}^{x}\sigma_{k+1}^{x} +\sigma_{k}^{y}\sigma_{k+1}^{y} +\Delta\sigma_{k}^{z}\sigma_{k+1}^{z}\right],
$$
where the sum is taken over the sites of a one-dimensional lattice and $\sigma^{x,y,z}$ are Pauli matrices.
Let us parametrize the anisotropy as $\Delta=\cos\pi\nu$. Then for the EFP we have \cite{KMST_gen-2002,KLNS-2002}
\be
    P(n) \sim  A n^{-\gamma}C^{-n^{2}},
\ee
as $n=2R\to \infty$,
where
\be
    C =\frac{\Gamma^{2}(1/4)}{\pi\sqrt{2\pi}} \exp\left\{-\int_{0}^{\infty}\frac{dt}{t}\,
    \frac{\sinh^{2}(t\nu) e^{-t}}{\cosh (2t\nu) \sinh (t)}\right\}
\ee
and the exponent $\gamma$ was conjectured in \cite{KLNS-2002} to be
\be
    \gamma = \frac{1}{12} +\frac{\nu^{2}}{3(1-\nu)}.
\ee



\end{document}